\documentclass[copyright,creativecommons]{eptcs}
\providecommand{\event}{AMCA-POP 2010} 

\providecommand{\firstpage}{1}

\usepackage{breakurl}             
\usepackage{graphicx}

\title{Modelling of Multi-Agent Systems: Experiences with Membrane Computing and Future Challenges}
\author{Petros Kefalas and Ioanna Stamatopoulou
\institute{Dept. of Computer Science CITY College, Thessaloniki, Greece\\
International Faculty of the University of Sheffield}
\email{\{kefalas, istamatopoulou\}@city.academic.gr}
}

\def\titlerunning{Modelling of Multi-Agent Systems: Experiences with Membrane Computing and Future Challenges}
\def\authorrunning{P. Kefalas \& I. Stamatopoulou}
\begin{document}
\maketitle

\begin{abstract}
Formal modelling of Multi-Agent Systems (MAS) is a challenging task due to high complexity, interaction, parallelism and continuous change of roles and organisation between agents. In this paper we record our research experience on formal modelling of MAS. We review our research throughout the last decade, by describing the problems we have encountered and the decisions we have made towards resolving them and providing solutions. Much of this work involved membrane computing and classes of P Systems, such as Tissue and Population P Systems, targeted to the modelling of MAS whose dynamic structure is a prominent characteristic. More particularly, social insects (such as colonies of ants, bees, etc.), biology inspired swarms and systems with emergent behaviour are indicative examples for which we developed formal MAS models. Here, we aim to review our work and disseminate our findings to fellow researchers who might face similar challenges and, furthermore, to discuss important issues for advancing research on the application of membrane computing in MAS modelling.
\end{abstract}

\section{Multi-Agent Systems and Formal Methods}

Software artefacts are characterised as agents if they can exhibit autonomous, reactive, proactive and social behaviour \cite{WoJe95}.
Autonomy is a property that allows agents to carry out their own thread of computation, without (much) intervention.
Reactivity is not classified as an intelligent behaviour, however, it is essential to provide immediate response to the percepts from the environment.
Sometimes, reactivity alone is more than enough to develop an emergent behaviour of a system \cite{Bro91}.
The operation of intelligent agents is driven by goals that are achieved through a sequence of actions planned. 
Such goal-oriented (proactive) behaviour often involves a rather complex deliberation process.
Finally, agents are able to communicate with other agents, a behaviour which leads to interaction between agents.

Multi-agent systems (MAS) consist of independent agents that can collaborate, negotiate, compete etc. towards the achievement of personal or shared goals.
MAS are rather complex, highly interactive, highly parallel and highly dynamic systems.
Agents play different roles in a MAS but they need to exchange and share information and knowledge in order to engage in a common problem solving activity.
This, apart from the need for certain communication and interaction protocols, requires an effective organisation between agents.
Organisation in a MAS, such as agent roles, communication structure, number of participating agents etc., is not static; it changes all the time throughout its operation.
These dynamics make MAS a challenging software development activity.
The more complex a MAS is, the more difficult the modelling process turns out to be and, in consequence, the less easy it is to ensure correctness at the modelling and implementation level.
Correctness implies that all desired safety properties are verified at the end of the modelling phase and that an appropriate testing technique is applied to prove that the implementation has been built in accordance to the verified model \cite{HoIp98}. 

In software engineering the term formal methods is used to classify mathematical approaches to all stages of software development.
The main arguments in favour of formal methods are rigour, expressiveness and the ability to reason.
The latter led to the promise of delivering correct software, i.e. software that is developed based on formal specifications and proofs (verified and tested), such that it performs in a desired manner under all circumstances.
There is a dispute whether formal methods have delivered what they promised, i.e. correctness, but no one can argue that research and practice have shown a considerable number of successes.

With MAS the issue of correctness is much more complicated, since MAS are open systems, often exhibiting unpredictable emergent behaviour, and organised around a rather complex structure, with agents that intensively communicate, continuously change roles, interact etc.
Therefore, formal modelling and verification that lead to implementation, testing, and simulation are challenging issues in MAS.

In our area of interest, formal modelling is particularly appealing as it raises many issues that cannot be tackled in a straightforward manner and leave many open challenges.
More specifically we have been investigating, among others, the suitability of classes of Membrane Computing systems \cite{pshandbook} as means of formal modelling of agents and MAS.
We have mainly focused our efforts to developing models for biological systems with emergent behaviour or biology inspired systems. 
In this paper we record our research experience on formal modelling for MAS.
We review the last decade work, by describing problems and their solutions.
This is aimed at disseminating our findings to fellow researchers who might face similar challenges. We also focus on important issues for advancing research on formal methods in MAS further.

\section{Case studies for Multi-Agent Systems}

During the last years we have been researching on the formal modelling of MAS. 
Before we start describing our cumulative experience, we will briefly summarise the kind of MAS we have been dealing with.

One class of MAS that was thought to be of particular interest was the biological systems and mostly systems of social insects.
Colonies of ants and bees as well as cell grouping fall within this category \cite{SSKE08, JRH04, KSSE09, KEHS05, GHK01}.
Both consist of relatively simple reactive agents that if left alone there is nothing much they can do.
Organised as colonies, however, with roles and direct or indirect communication exhibit an emergent intelligent behaviour.
For example, Pharaoh ants \cite{SSKE08, JRH04}, apart from the pheromone trail they produce to be used for foraging, they have a very effective way of organising their work within the society as well as help each other to survive by exchanging food.
Japanese bees \cite{KSSE09} contribute as individuals to increase the temperature in a hive and burn the attacking hornet.
Foraging bees communicate the destination of the food source by the famous dance of the returning bee \cite{GHK01}.
Similar emergent behaviour but with less direct communication among agents can be achieved in flocks, schools and herds \cite{SGK05, PEC08}.

The other class of MAS that we have been modelling are those characterised as biology-inspired.
Such a system is NASA ANTS \cite{CMNMRB00}, a MAS which deploys unmanned spacecrafts with a variety of specialisations to explore asteroids.
More particularly, spacecrafts are organised in groups, each one consisting of a leader, multiple workers, and one or more messengers. Workers are responsible for gathering measurements, messengers for coordinating communication among all involved spacecrafts, and leaders for gathering measurements from workers, setting goals and coordinating group formations.
We chose ANTS because it provides a good testbed for applying formal methods \cite{RVTRH04}.
The important feature of ANTS is that there is a rather strict organisation which however is not affected even though there might be a large number of spacecrafts out of order or destroyed \cite{SKG07e}.
Robustness is a property of all swarm intelligence systems.

\section{Key issues in the modelling of agents and multi-agent systems}

Individual agents operate based on the following: 
\begin{itemize}
\item they perceive their environment by receiving stimuli as input which they filter and accept for further processing;
\item they receive messages from other agents;
\item they update their beliefs based on both the percepts as well as the information encoded in the received messages, by revising their temporary knowledge about the environment and others;
\item they react based on a specific set of rules that describe individual behaviours;
\item they engage in a deliberation process, which allows them to revise their goals and plans, and decide what is the next action to be performed;
\item they compile and send messages to other agents;
\item they act and the effects of their action appear in the environment.
\end{itemize}

Not all the above are present in every agent. For example, reactive agents do not deliberate, while ``smarter'' proactive agents do. Also, communication between simple biological agents is rather primitive and mostly done through the environment, in contrast to more elaborated direct communication that may follow a strict protocol. Therefore, in order to create a model of an agent, one would require:
\begin{itemize}
\item non-trivial data structures, e.g. set of \textit{n}-tuples, sequences, lists, terms, with a set of their corresponding operations, to represent beliefs, goals, plans, messages, percepts etc.; 
\item means of encoding rules that express reactive behaviour, which can also be arranged in a strict order, e.g. avoid collisions, follow trail, forage;
\item means of encoding the functionality that corresponds to their proactive behaviour (if such must be present), such as revision of goals and plan generation;
\item representation of the internal state of the agent.
\end{itemize}

In a multi-agent system:
\begin{itemize}
\item each agent operates in parallel with others;
\item the mode of interaction imposes the way in which agents exchange messages;
\item the roles and organisation define the structure of the communication flow;
\item new agents may come into play while other cease to exist.
\end{itemize}

At MAS level, modelling would require:
\begin{itemize}
\item ways to deal with exchange of messages between agents, either direct or indirect through the environment, deterministic broadcast, peer-to-peer or non-deterministic etc.;
\item a way to express the interaction with the environment, that is, perception and action;
\item a method for expressing the asynchronous computation of individuals;
\item the addition and removal of agent instances ``on the fly'';
\item means for structuring and restructuring the organisation ``on the fly'' (structure mutation).
\end{itemize}

If the modelling method used is formal, then there are a number of consequences that accompany this choice.
First of all, formal reasoning on the model can be performed.
This can be through formal verification, either proofs or model checking. 
Formal verification \cite{EmCl81} will check whether desired properties of individual agents, or ideally of the whole system, stand.
This is rather crucial before someone proceeds with implementation.
Secondly, one can employ formal testing techniques.
A set of test cases can be produced from the model to check whether the implementation is correct with respect to the model \cite{IpHo97}. 
Thirdly, if the formal method is accompanied by tools, prototype animation or simulation may be possible \cite{flame}.
This would facilitate the identification of misconceptions in the model which can then be fixed before proceeding to the implementation.
Finally, a set of refinement transformations could safely lead to an implementation of the system that matches the original specification and model.

\begin{table}[h]
\begin{center}
\caption{Comparison of features of CXS, tPS, PCol and PPS with respect to modelling.}
\label{cxm-vs-pps}
\begin{tabular}{lllll}
\hline\noalign{\smallskip}
\textbf{Modelling feature} & \textbf{CXS} & \textbf{tPS} & \textbf{PCol} & \textbf{PPS}  \\
\noalign{\smallskip}\hline\noalign{\smallskip}
\textbf{Individual Agents} & & & & \\
\noalign{\smallskip}\hline\noalign{\smallskip}
Agent internal state representation & $\surd$ & $\times$ & $\times$ & $\times$\\
Rules to describe reactive behaviour & $\surd$ & $\times$ & $\surd$ & $\times$\\
Rules to describe proactive behaviour & $\times$ & $\times$ & $\times$ & $\times$\\
Non-trivial data structures for beliefs, goals, messages, stimuli etc & $\surd$ & $\times$ & $\times$ & $\times$\\
Formal verification of individual agents & $\surd$ & $\times$ & $\times$ & $\times$\\
Test case generation for individual agents & $\surd$ & $\times$ & $\times$ & $\times$\\
\noalign{\smallskip}\hline\noalign{\smallskip}
\textbf{Communication} & & & & \\
\noalign{\smallskip}\hline\noalign{\smallskip}
Direct communication and message exchange & $\surd$ & $\surd$ & $\times$ & $\times$\\
Non-deterministic communication & $\times$ & $\surd$ & $\surd$ & $\surd$ \\
Indirect communication through the environment & $\times$ & $\surd$ & $\surd$ & $\surd$ \\
Environmental stimuli (input) & $\surd$ & $\surd$ & $\surd$ & $\surd$ \\
Perception & $\times$ & $\times$ & $\times$ & $\times$ \\
\noalign{\smallskip}\hline\noalign{\smallskip}
\textbf{MAS Structure} & & & & \\
\noalign{\smallskip}\hline\noalign{\smallskip}
Definition of agent roles & $\surd$ & $\surd$ & $\times$ & $\surd$ \\
Addition of agent instances on the fly & $\times$ & $\times$ & $\times$ & $\surd$ \\
Removal of agent instances on the fly & $\times$ & $\times$ & $\times$ & $\surd$ \\
Communication network restructuring & $\times$ & $\times$ & $\times$ & $\surd$ \\
\noalign{\smallskip}\hline\noalign{\smallskip}
\textbf{MAS Operation} & & & & \\
\noalign{\smallskip}\hline\noalign{\smallskip}
Maximal parallelism & $\surd$ & $\surd$ & $\surd$ & $\surd$ \\
Arbitrary parallelism & $\surd$ & $\surd$ & $\surd$ & $\surd$ \\
MAS verification and testing & $\times$ & $\times$ & $\times$ & $\times$\\
Tool support & $\surd$ & $\times$ & $\times$ & $\surd$ \\
\noalign{\smallskip}\hline\noalign{\smallskip}
\textbf{Environment} & & & & \\
\noalign{\smallskip}\hline\noalign{\smallskip}
Modelling of the environment & $\times$ & $\surd$ & $\surd$ & $\surd$ \\
\noalign{\smallskip}\hline
\end{tabular}
\end{center}
\end{table}

In our experimentation, we tried out two types of formal methods, namely state-based methods and membrane computing.
For the modelling process we have investigated a number of instances of those methods, such as X-Machines (XM) \cite{Eil74, Hol88} and Communicating X-Machines (CSX) \cite{KEK01} for the former, as well as tissue P systems (tPS) \cite{MvPPRp03}, P Colonies (PCol) \cite{KKP04} and Population P Systems \cite{BeGh04} with active cells (PPS) for the latter. Table \ref{cxm-vs-pps} shows a comparison between all methods, as to whether they can satisfy ---directly, not through implicit means--- the key issues in MAS modelling mentioned above.
The comparison refers to the most widely used definitions of the models.
There are actually numerous extensions that one way or another try to enhance the existing definitions with additional features. 
It should also be noted that X-Machines are not included in the comparison as an X-Machine model may only represent a single agent whereas in the table we compare formalisms that may be used for the modelling of MAS.

\section{Methods employed for modelling of MAS}

After carefully considering the aforementioned alternatives and based on the comparison presented above we selected to work with Communicating X-Machines and Population P Systems with active cells, and attempted a number of modelling approached for all the MAS mentioned above, e.g. biological cells, flocks, ants, Pharaoh ants, foraging bees, Japanese bees, and NASA ANTS. The reason for selecting CXMs is due its advantages in regards to the modelling of an individual agent's behaviour. Out of the three membrane computing formalisms we selected PPSs with active cells due to the fact that they best support operations on the MAS structure, such as addition and removal of agents as well as communication network restructuring.

A Communicating X-Machine System $CXMS = (XM_i, R),\ 1\leq i \leq n$ is a collection of $n$ X-machines $XM_i$ able to communicate through channels, as they defined in the communication relation $R$.
More particularly, an $XM_i$ is a deterministic stream X-machine \cite{HoIp98} defined as follows:
$$X = (\Sigma,\: \Gamma,\: Q,\: M,\: \Phi,\: F,\: q_0,\: m_0)$$
where:
\begin{itemize}
\item $\Sigma$ and $\Gamma$ are the input and output alphabets, respectively.
\item $Q$ is the finite set of states.
\item $M$ is the (possibly) infinite set called memory.
\item $\Phi$ is a set of partial functions $\varphi$; each such function maps an input and a memory value to an output and a possibly different memory value, $\varphi: \Sigma \times M \rightarrow \Gamma \times M$. 
\item $F$ is the next state partial function, $F: Q\times\varphi \rightarrow Q$, which given a state and a function from the type $\Phi$ determines the next state. $F$ is often described as a state transition diagram.
\item $q_0$ and $m_0$ are the initial state and initial memory respectively.
\end{itemize}

\begin{figure}[ht]
	\centering
	\scalebox{1}{\includegraphics{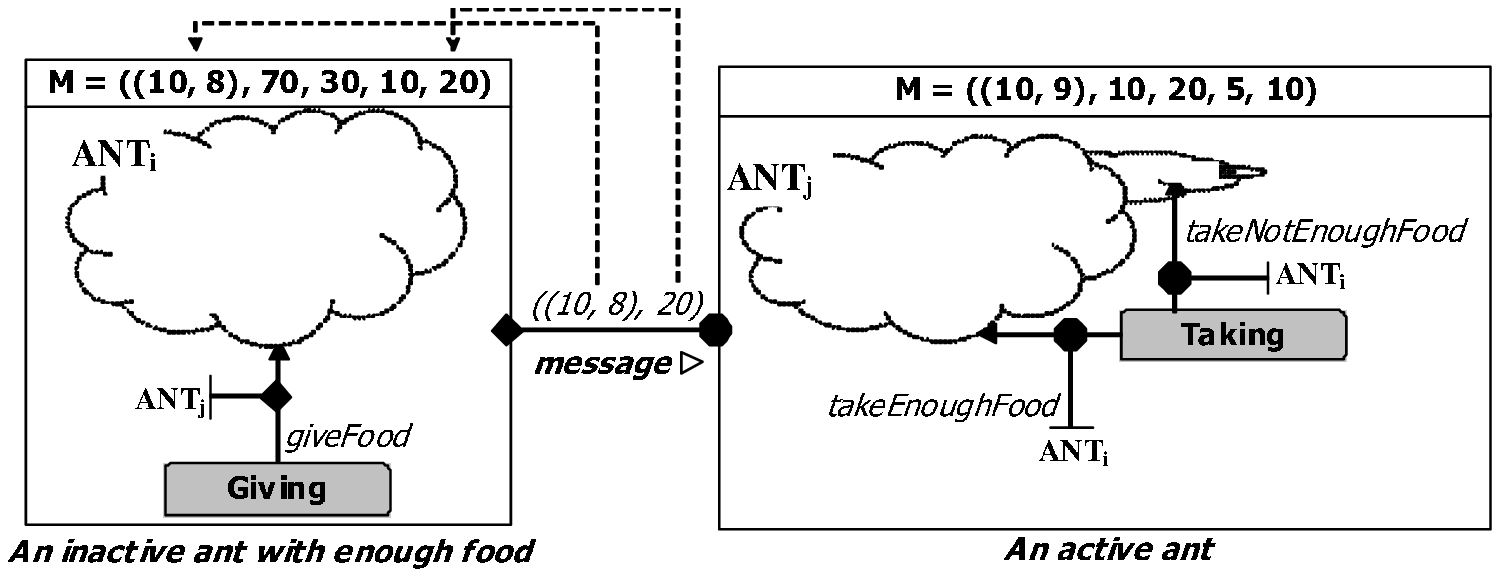}}
	\caption{An example of a Communicating X-machine modelling the exchange of food between two ants.}
	\label{comant}
\end{figure} 

Note that the definition of an $XM_i$ that belongs to a communicating system is slightly different, but for reasons of exposition it is not appropriate to elaborate here.

As an example consider the case of the Pharaoh ants. A more complete model of this case study may be found in \cite{Sta08} but it is partially included here for demonstration purposes. 

The ants spend much of their in-nest time doing nothing, thus staying inactive. An ant may become active when its food reserves drop below a defined minimum threshold.

The assumptions that are made in this study are the following:
\begin{itemize}
\item the colony only consists of workers;
\item the nest, in which the colony is situated, is a rectangular environment (2D grid);
\item the ants are either inactive or move around looking for food. If no food is found, they go outside the nest to forage and identify (new) locations for food;
\item when two ants meet they might share food, if one is actively searching for food and the other has food reserves;
\item the ants go out to forage when they do not have sufficient food reserves (according to the food quantity threshold), no food source is identified and a pheromone trail leading to an exit of the nest is discovered;
\item ants that are outside may enter the nest at any time.
\end{itemize}

\begin{figure}[ht]
	\centering
	\scalebox{1}{\includegraphics{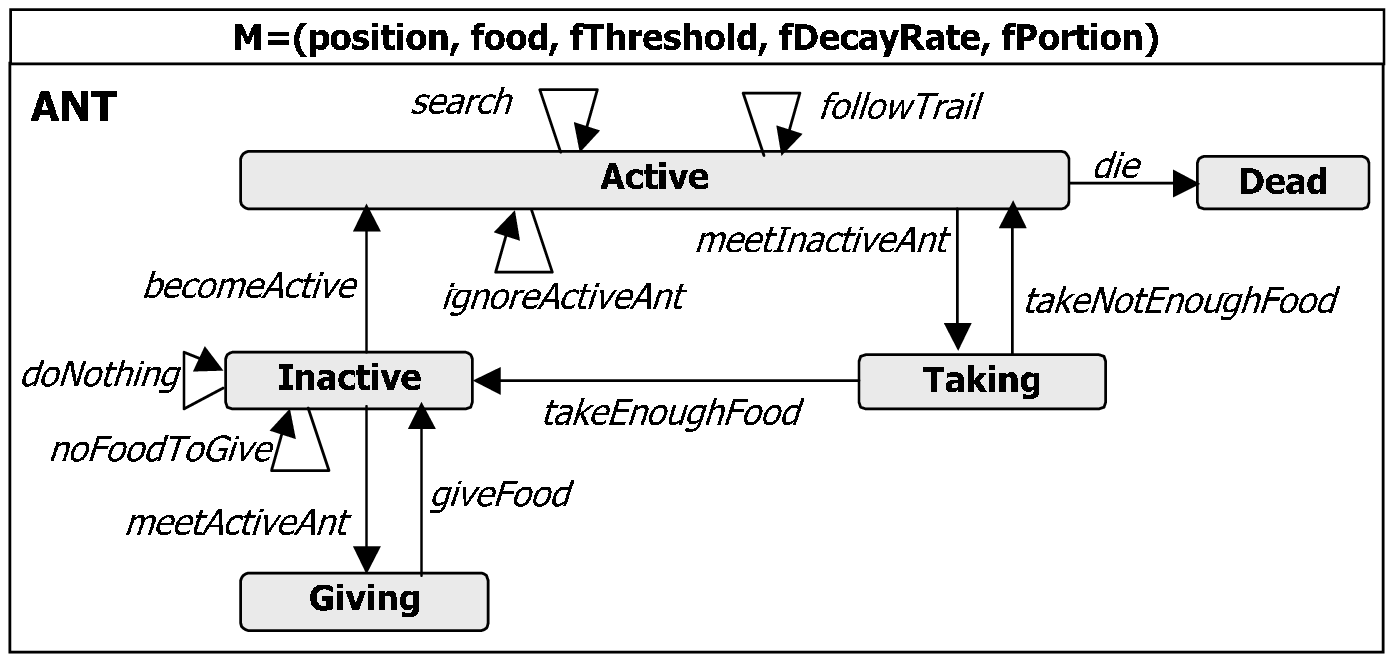}}
	\caption{An example of an X-machine modelling a Pharaoh ant.}
	\label{ant}
\end{figure} 

An example of a Communicating X-machine in regards to the above description is depicted in Fig. \ref{comant} and shows how two ants communicate by sharing food. The inactive ant's function $giveFood$ sends as output the food amount it is willing to share to be received as input by the $takeEnoughFood$ function of the active ant. Fig. \ref{ant} shows the state transition diagram of the $XM$ model of one individual Pharaoh ant.

A Population P System with Active Cells \cite{BeGh04} is defined as a construct:
$$P=(V,\: K,\: \gamma,\: \alpha,\: w_E,\: C_1,\: C_2,\: \ldots,\: C_n,\: R)$$
where:
\begin{itemize}
\item $V$ is a finite alphabet of symbols called objects;
\item $K$ is a finite alphabet of symbols, which define different types for the cells;
\item $\alpha$ is a finite set of bond-making rules of the general form $(t,x_1;x_2,p)$, with $x_1,x_2 \in V^*$, and $t,p \in K$;
\item $\gamma=(\{1,2,\ldots\,n\},A)$, with $A \subseteq \{\{i,j\}\, | \,1 \leq i \neq j\leq n\}$, is a finite undirected graph;
\item $w_{\mathrm{E}} \in V^*$ is a finite multiset of objects initially assigned to the environment;
\item $C_i=(w_i,t_i)$, for each $1\leq i \leq n$, with $w_i \in V^*$ being a finite multiset of objects, and $t_i \in K$ the type of cell $i$;
\item $R$ is a finite set of:
	\begin{itemize}
	\item communication rules of the form $r:(\,a\,;\,b,in\,)_t$, $r:(\,a\,;\,b,enter\,)_t$, $r:(\,b,exit\,)_t$, for $a\in V \cup\{\lambda\}$, $b \in V$, $t \in K$, which allow the moving of objects between neighbouring cells or a cell and the environment according to the cell type and the existing bonds among the cells;
	\item object transformation rules of the form $r:(\,a\rightarrow b\,)_t$, for $a \in V$, $b \in V^+$, $t \in K$, meaning that an object $a$ is replaced by an object $b$ within a cell of type $t$;
	\item cell differentiation rules of the form $r:(\,a\,)_t \rightarrow (\,b\,)_p$, with $a,b \in V$, $t,p \in K$ meaning that consumption of an object $a$ inside a cell of type $t$ changes the cell, making it become of type $p$. All existing objects remain the same besides $a$ which is replaced by $b$;
	\item cell division rules of the form $r:(\,a\,)_t \rightarrow (\,b\,)_t\,(\,c\,)_t$, with $a,b,c \in V$, $t\in K$ meaning that a cell of type $t$ containing an object $a$ is divided into two cells of the same type. One of the new cell has $a$ replaced by $b$ while the other by $c$;
	\item cell death rules of the form $r:(\,a\,)_t \rightarrow \dagger$, with $a \in V$, $t\in K$ meaning that an object $a$ inside a cell of type $t$ causes the removal of the cell from the system.
	\end{itemize} 
\end{itemize}

\begin{figure}[ht]
	\centering
	\scalebox{0.9}{\includegraphics{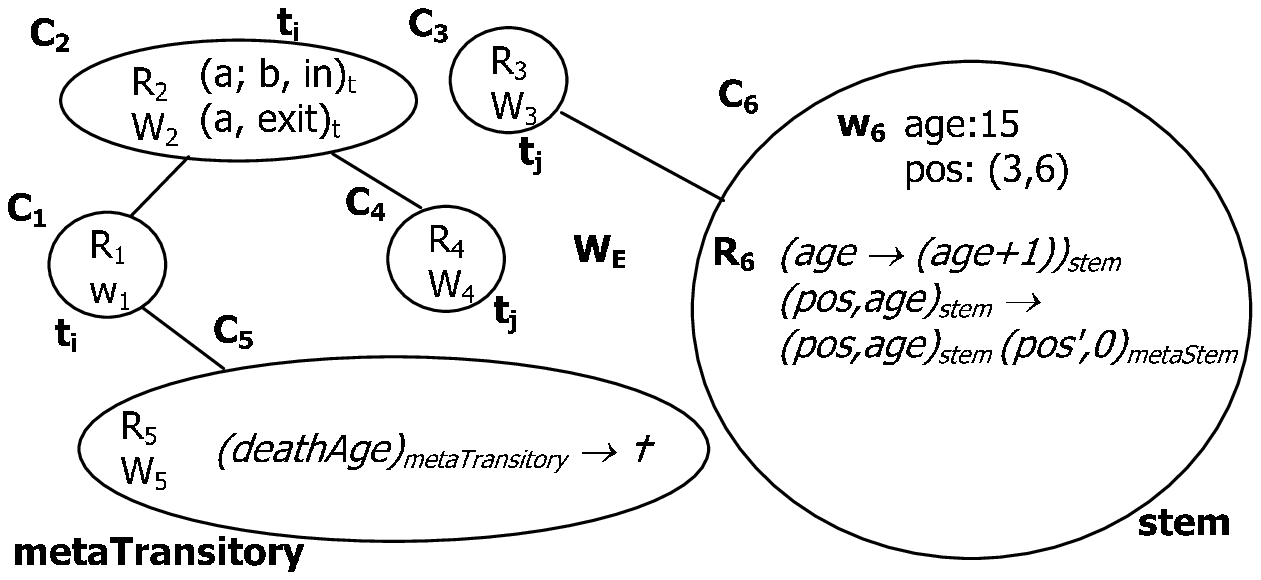}}
	\caption{An example of a Population P System; $R_i$: set of rules related to cell $C_i$; $w_i$: multiset of objects associated to cell $C_i$.}
	\label{newfig}
\end{figure}

An example of a Population P System modelling tumour growth is depicted in Fig. \ref{newfig} (model description borrowed from the NetLogo models library \cite{netlogo}; a complete system definition using PPS may be found in \cite{Sta08}). 

A tumour consists of two kinds of cells: stem and transitory cells. It is a stem cell that is required for the formation of the tumour to begin. At each time unit all cells divide thus doubling the size of the tumour. Stem cells may divide in two ways: (a) \textit{asymmetrically}, thus breeding a transitory cell that moves outward, and (b) \textit{symmetrically}, breeding another stem cell which also moves outward and settles in another location thus creating a metastasis of the original tumour. In effect, a stem cell never dies as during division one of the two daughter cells is always a stem cell.

Transitory cells divide only symmetrically up to a certain age; after that age they mature and eventually die. Finally, transitory cells that have originated from a metastatic stem cell, called metatransitory, as well as all their offsprings die younger.

Each cell type has its own objects and rules. For instance, in Fig. \ref{newfig} $C_6$ is a $stem$ cell with two rules according to the given example: a transformation rule that represents the cell's ageing by increasing it by 1, and a division rule that divides the cell in two creating a new cell in position $pos'$ with an age of 0.

\section{Synthesised Methods and Transformations}

It is apparent that the two types of methods (state-based and membrane) are complementary to MAS modelling needs, something which led us to some kind of synthesis of the two.
We attempted a potential integration of the two as an instance of the $OPERAS$ framework \cite{Sta08}.

The $OPERAS$ formal framework for defining a dynamic multi-agent system, is defined by the 6-tuple $(O, P, E, R, A, S)$ where:
\begin{itemize}
	\item $O$ contains the reconfiguration operations (or rules, e.g. of the general form $condition \Rightarrow action$). Each operation involves the application of one or more of the operators that create or remove a communication channel between agents or introduce/remove an agent in/from the system;
	\item $P$ is the distributed union of the percepts of all the types of agents involved in the system;
	\item the communication relation $R: A \times A$ with $(A_i, A_j) \in R$, $A_i, A_j \in A$ conveys the information that agents $A_i, A_j$ can communicate by exchanging messages;
	\item $E$ is a model of the environment;
	\item $A$ is the set of agent instances $A = \{A_1, \ldots A_n\}$ where $A_i$ is an agent instance defined in terms of (a) its individual behaviour, and (b) its local structural mutation mechanism for reconfiguring the system structure in its proximity;
	\item the set $S = \{(Behaviour_t, StrMut_t)\: |\: t \in Types\}$ holds the definitions of agent types ($Types$ being a set of identifiers of the types of agents).
\end{itemize}

\begin{figure}[ht]
	\centering
	\scalebox{1}{\includegraphics{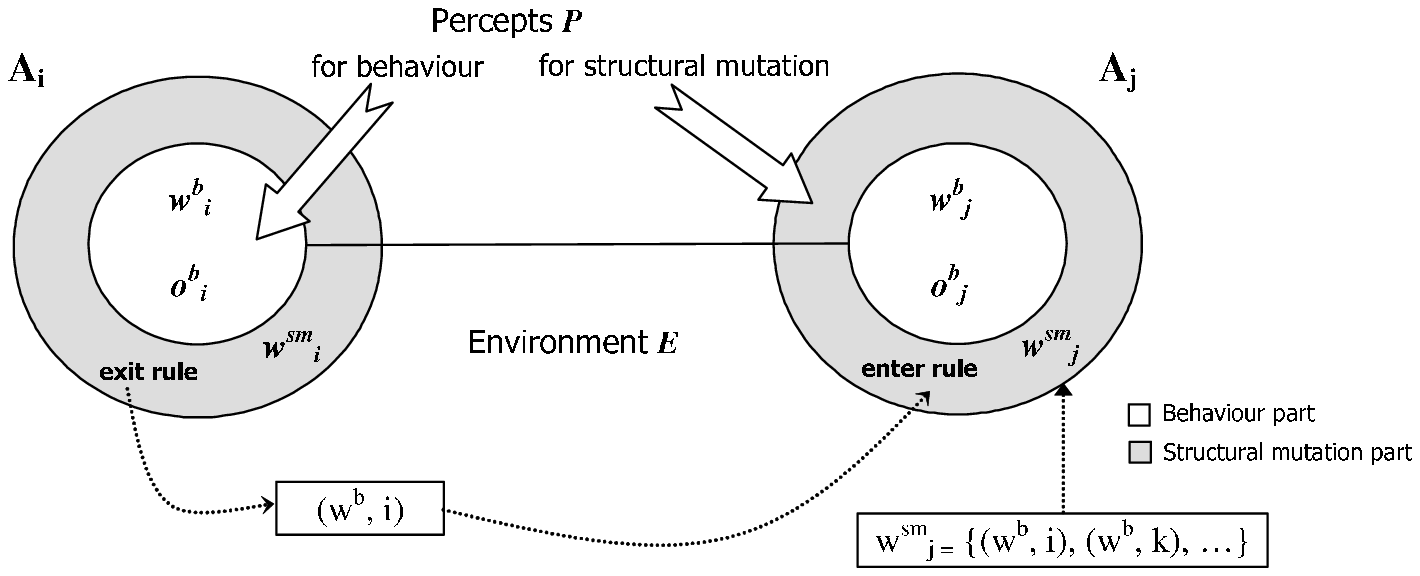}}
	\caption{An abstract example of an $OPERAS_{CC}$ model that uses Population P System concepts for the representation of both the agent's behaviour and structure dynamics.}
	\label{operasCC}
\end{figure}

In OPERAS the behaviour of an agent can be modelled separately from its control. 
In principle, this means that one can employ two different formal methods for each, thus taking advantage of both state-based models and membrane computing ideas.
It is therefore implied that there are several options which could instantiate OPERAS into concrete modelling methods. 
As mentioned above, we have long experimented with Communicating X-machines and Population P Systems with active cells, thus resulting into hybrid models such as $OPERAS_{CC}$ \cite{SKG07c} and $OPERAS_{XC}$ \cite{SKG07e}.
The former uses PPS features for both modelling the dynamics and the behaviour of the agents, as is abstractly depicted in Fig. \ref{operasCC}.

\begin{figure}[ht]
	\centering
	\scalebox{0.7}{\includegraphics{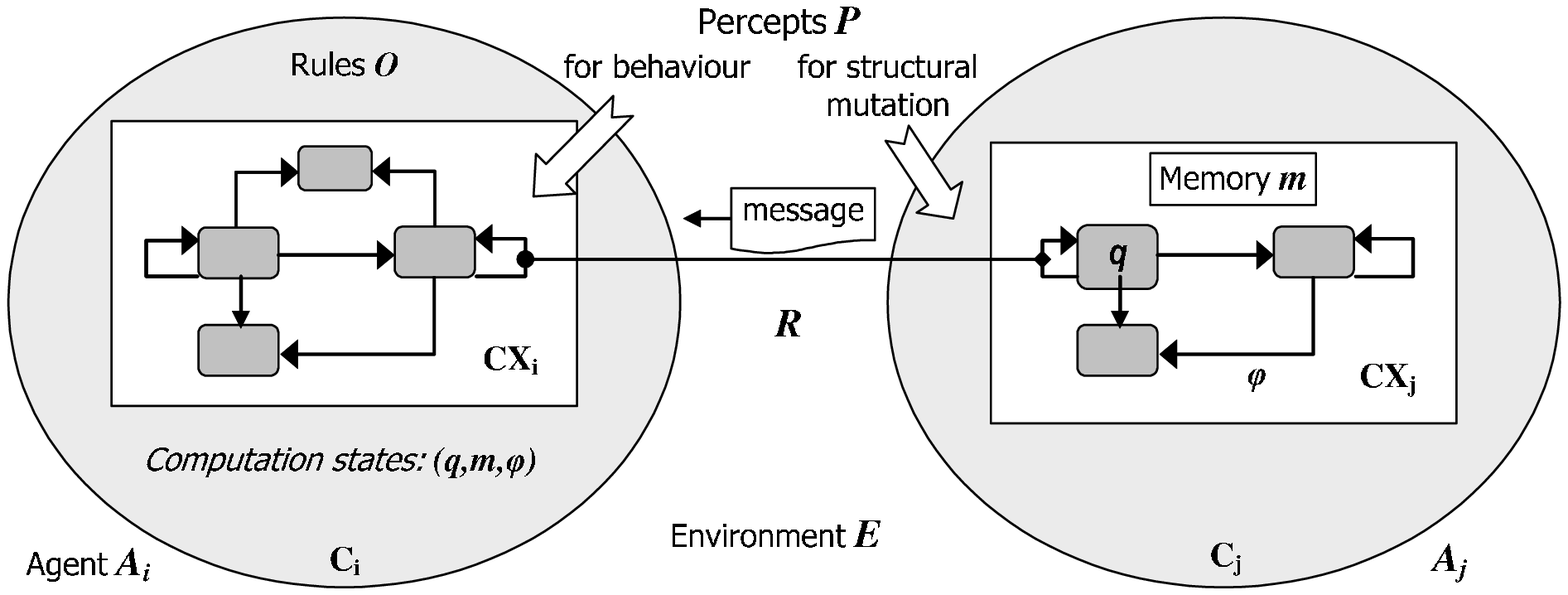}}
	\caption{An abstract example of an $OPERAS_{XC}$ model that uses state machines for the behaviour and PPS-like rules for the organisation of the system.}
	\label{operasXC}
\end{figure}

$OPERAS_{XC}$ on the other hand is a combination that uses state machines for the behaviour and PPS-like rules for the organisation of the system, as is abstractly depicted in Fig. \ref{operasXC}.
This gave us the opportunity to combine the advantages that XMs have in terms of modelling the behaviour of an agent with the advantages that PPSs have in terms of defining the control over the structure of the system.
Also, the computation is driven by either method, which leads to a variety of interesting overall MAS computation. For a complete case study modelled using $OPERAS_{XC}$ the interested reader is referred to \cite{Sta08}.

We have also attempted a number of transformations between Communicating X-Machines and P Systems and vice-versa \cite{KEHG03, KSSE09}. 
Others in the P system research community showed more interest in other transformations, such as from P Systems to Petri nets \cite{KlKo07}, process algebra \cite{CiAm07}, cellular automata \cite{CoFr07} etc.
These have not only demonstrated the equivalence of these methods but created a temptation to try out several techniques known for one method to the other.

In the development of the MAS models, two notations have been created, namely X-Machine Definition Language (XMDL) \cite{KaKe00} and Population P System Definition Language (PPSDL) \cite{SKEG05}.
The accompanied tools assisted us in understanding in depth the overall behaviour of the developed models through textual animation.

\section{Discussion and Open Issues}

There are several open issues that are left for further research on MAS development, namely formal modelling, verification, testing and tool support.

With respect to modelling, the effort to define a special class of P Systems should continue.
The new class must employ all features that will facilitate agent and multi-agent system modelling.
The computation within cells must be enhanced so that it can offer some complex reasoning required for goal-oriented agents.
Such an attempt may inevitably restrict maximal parallelism for the sake of correctness of the behaviour of agents as well as of the whole system.
An initial first attempt can be found at \cite{KeSt10}.
There is also room for development of the perception of the environment, since inserting an object into the cell may not be enough in realistic systems.
MAS researchers would also like to see built-in features towards formal modelling of interaction between cells. 

So far, there have not been useful results reported for verification of MAS models.
Although verification of individual agents is possible, verification of the complete MAS seems like an insurmountable obstacle.
The obvious reason is the combinatorial explosion problem due to the number of interactions that increase the state space in an exponential manner.
It would be worth investigating whether there exist methods that work under specific harmless assumptions that could omit non-safety properties and reduce the search space.
Finally, model checking techniques for P systems is an interesting area open for research developments.
A direction towards model checking would be the automatic or semi-automatic translation of models into code, for model checkers such as SPIN \cite{spin} or SMV \cite{smv}.

If formal verification seems hard to achieve, informal techniques, such as simulation have a lot to offer \cite{psweb_sw}.
Suitable and correct transformations of formal models could lead to executable models that simulate MAS.
In turn, simulation can facilitate the discovery of erroneous situations or undesired behaviour of the system.
For numerous types of MAS, biology inspired included, visual animation is highly desirable \cite{netlogo, SSKE08}.

Finally, although there exist testing techniques that can identify all faults in the implementation of an individual agent developed based on state-based models, there is little work done towards the testing of membrane systems \cite{IpGh09}.
Inevitably, it would appear to be a challenge to invent equivalent techniques for the whole multi-agent system.

\section{Conclusion}

We have presented a review on the use of Population P Systems with active cells in Multi-Agent System modelling. 
In the process of developing formal models of MAS, we discovered a number of challenging issues that could partly be addressed by state based models and partly by membrane computing models.
These characteristics were pinned down together with the available features of various methods that could make modelling possible.
The synthesized solution gives space to a number of challenges, such as verification, testing and simulation.
With this review, we attempted to disseminate our findings, initialise discussion that will set up directions of future research.

\bibliographystyle{eptcs} 
\bibliography{biblio}

\end{document}